\documentclass[twocolumn,showpacs,prl,amsmath,amssymb]{revtex4}

\usepackage{graphicx}
\usepackage{dcolumn}
\usepackage{bm}

\begin{document}

\title{Model of two-fluid reconnection}

\author{Leonid M. Malyshkin}
\email{leonmal@uchicago.edu}
\affiliation{Department of Astronomy \& Astrophysics,
University of Chicago, 5640 S. Ellis Ave., Chicago, IL 60637}
\date{\today}

\begin{abstract}
A theoretical model of quasi-stationary, two-dimensional
magnetic reconnection is presented in the framework of incompressible 
two-fluid magnetohydrodynamics (MHD). 
The results are compared with recent numerical simulations and experiment.
\end{abstract}  


\maketitle


\section{\label{INTRODUCTION}
Introduction
}
Magnetic reconnection is a fundamental physical process of topological 
rearrangement of magnetic field lines in magnetized plasmas during which
magnetic energy is converted into kinetic and thermal energy. Reconnection
is of particular importance in the solar atmosphere, the Earth's 
magnetosphere, and in laboratory 
plasmas~\citep{biskamp_2000,drake_2006,zweibel_2009,yamada_2009}. 
In hot plasmas, because of the low electrical resistivity, 
magnetic reconnection due to resistive dissipation alone is very slow. 
As a result, a simple single-fluid MHD description of the plasma is 
generally believed to be insufficient for the theoretical explanation of fast 
reconnection events. 
Instead, a two-fluid MHD approach has been frequently used in 
recent studies of fast reconnection~\citep{biskamp_2000,
drake_2006,zweibel_2009,yamada_2009}. Most these studies have been 
numerical and experimental, while an ultimate goal of construction 
a comprehensive theoretical model of two-fluid reconnection has not 
yet been achieved. In this Letter we present a model 
of two-fluid magnetic reconnection, which serves this goal. 

\phantom{o}\hspace{-1.5cm}\phantom{o}


\section{\label{EQUATIONS}
Two-fluid MHD equations
}
We use physical units in which the speed of light $c$ and four times 
$\pi$ are replaced by unity, $c=1$ and $4\pi=1$. 
To rewrite our equations in the Gaussian centimeter-gram-second (CGS) 
units, one needs to make the following substitutions: magnetic field 
${\bf B}\rightarrow {\bf B}/\sqrt{4\pi}$, electric field 
${\bf E}\rightarrow c{\bf E}/\sqrt{4\pi}$, electric current 
${\bf j}\rightarrow\sqrt{4\pi}\,{\bf j}/c$, electrical resistivity 
$\eta\rightarrow\eta c^2/4\pi$, the proton charge 
$e\rightarrow\sqrt{4\pi}\,e/c$.

We consider an incompressible, non-relativistic and quasi-neutral 
plasma, composed of electrons and protons. Using standard notation, 
the equations of motion for the electrons and protons 
are~\citep{biskamp_2000,sturrock_1994} 
\begin{eqnarray}
nm_e\!\left[\partial_t{\bf u}^e
\!+\!({\bf u}^e{\bf\nabla}){\bf u}^e\right]\!=\!-{\bf\nabla}P_e
-ne({\bf E}+{\bf u}^e\!\times\!{\bf B})-{\bf K},&&
\label{MOTION_E}
\\
nm_p\!\left[\partial_t{\bf u}^p
\!+\!({\bf u}^p{\bf\nabla}){\bf u}^p\right]\!=\!-{\bf\nabla}P_p
+ne({\bf E}+{\bf u}^p\!\times\!{\bf B})+{\bf K},&&
\label{MOTION_P}
\end{eqnarray}
where $n$ is the (constant) number density, and the subscripts and 
superscripts $e$ and $p$ refer to electrons and protons respectively.  
Here, ${\bf K}$ is the resistive frictional force due to electron-proton
collisions that can be approximated as 
${\bf K}=n^2e^2\eta({\bf u}^e-{\bf u}^p)=-ne\eta{\bf j}$, where 
$\eta$ is the electrical resistivity~\citep{biskamp_2000,sturrock_1994}. 
For simplicity, we neglect proton-proton and 
electron-electron collisions and the corresponding viscous forces. 
We also introduce the electric current ${\bf j}=ne({\bf u}^p-{\bf u}^e)$ 
and the plasma velocity ${\bf V}=n(m_p{\bf u}^p+m_e{\bf u}^e)/\rho$, 
where $\rho=n(m_p+m_e)={\rm const}$ is the plasma density. 
Taking into account $m_e\ll m_p$, we find ${\bf u}^p={\bf V}+m_e{\bf j}/nem_p$ 
and ${\bf u}^e={\bf V}-{\bf j}/ne$. We substitute these expressions and 
${\bf K}=-ne\eta{\bf j}$ into Eqs.~(\ref{MOTION_E}) and~(\ref{MOTION_P}). 
We also substitute the electric field ${\bf E}$ from Eq.~(\ref{MOTION_E}) 
into Eq.~(\ref{MOTION_P}). We obtain
\begin{eqnarray}
&&{\bf E}=
\eta{\bf j}-{\bf V}\!\times\!{\bf B}+{\bf j}\!\times\!{\bf B}/ne
-[{\bf\nabla}P_e-(d_e^2/d_p^2){\bf\nabla}P_p]/ne
\nonumber
\\
&&\qquad
{}+d_e^2\left[\partial_t{\bf j}+({\bf V}{\bf\nabla}){\bf j}
+({\bf j}{\bf\nabla}){\bf V}
-(1/ne)({\bf j}{\bf\nabla}){\bf j}\right],
\label{OHMS_LAW}
\\
&&nm_p\left[\partial_t{\bf V}+({\bf V}{\bf\nabla}){\bf V}\right]=
-{\bf\nabla}P+{\bf j}\times{\bf B}-d_e^2({\bf j}{\bf\nabla}){\bf j},
\label{MOTION_LAW}
\end{eqnarray}
where $P=P_e+P_p$ is the total pressure, 
$d_e=(m_e/ne^2)^{1/2}$ and $d_p=(m_p/ne^2)^{1/2}$ 
are the electron and proton inertial lengths. 
Eq.~(\ref{OHMS_LAW}) is the generalized Ohm's law
describing the motion of the electrons. 
Eq.~(\ref{MOTION_LAW}) is the plasma momentum equation 
describing the motion of the protons. We note that the electron 
inertia terms, proportional to $d_e^2$, enter both Ohm's 
law and the momentum equation. Although these terms are essential 
for fast two-fluid reconnection (as we shall see presently), they
have been frequently neglected in the momentum 
equation before. We also note that ${\bf\nabla}\cdot{\bf V}=0$ and 
${\bf\nabla}\cdot{\bf j}=0$ for incompressible and non-relativistic 
plasmas.

\begin{figure}[t]
\vspace{2.7truecm}
\includegraphics{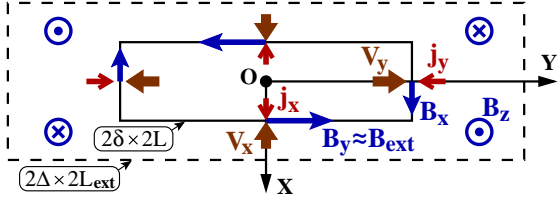}
\caption{Geometrical configuration of the reconnection layer.
}
\label{FIGURE_LAYER}
\end{figure}

We consider two-fluid magnetic reconnection in the classical 
Sweet-Parker-Petschek geometry, shown in
Fig.~\ref{FIGURE_LAYER}. The reconnection layer is in the $x$-$y$ 
plane with the $x$- and $y$-axes perpendicular to and along the 
layer respectively and all $z$ derivatives are zero.
The thickness of the reconnection current layer is $2\delta$, 
which is defined in terms of the out-of-plane current ($j_z$) 
profile across the layer~\footnote{$\delta$ can be 
formally defined by fitting the Harris sheet profile 
$(B_{ext}/\delta)cosh^{-2}(x/\delta)$ to function $j_z(x,y=0)$.
}. 
It can be shown that $2\delta$ turns out to be also the thickness 
of the electron layer, where the electrons are decoupled from the field 
lines. The length of the electron (current) layer is defined as $2L$. 
The proton layer, where the protons are decoupled from the field lines, 
has thickness $2\Delta$ and length $2L_{ext}$, which can be much larger 
than $2\delta$ and $2L$ respectively. The values of the reconnecting field 
outside the electron layer (at $x\approx\delta$) and outside the proton
layer (at $x\approx\Delta$) are about the same, $B_y\approx B_{ext}$ 
up to a factor of order unity. 
The out-of-plane field $B_z$ is assumed to have a quadrupole 
structure 
(see Fig.~\ref{FIGURE_LAYER})~\citep{drake_2006,yamada_2009,zweibel_2009}. 
Also, the reconnection layer is assumed to have a point symmetry with 
respect to its geometric center~$O$ (see Fig.~\ref{FIGURE_LAYER}) 
and reflection symmetries with respect to the $x$- and $y$-axes. 
Thus, 
$V_x(\pm x,\mp y)=\pm V_x(x,y)$, $V_y(\pm x,\mp y)=\mp V_y(x,y)$,
$V_z(\pm x,\mp y)=V_z(x,y)$,
$B_x(\pm x,\mp y)=\mp B_x(x,y)$, $B_y(\pm x,\mp y)=\pm B_y(x,y)$,
$B_z(\pm x,\mp y)=-B_z(x,y)$,
$j_x(\pm x,\mp y)=\pm j_x(x,y)$, $j_y(\pm x,\mp y)=\mp j_y(x,y)$
and $j_z(\pm x,\mp y)=j_z(x,y)$. The derivations below extensively 
exploit these symmetries and are similar 
to~\citep{malyshkin_2005,malyshkin_2008}.

\section{\label{SOLUTION}
Solution for two-fluid reconnection
}
We make the following assumptions for the reconnection process. 
First, $\eta$ is constant and small. Second, the reconnection process 
is quasi-stationary, so that we can neglect time derivatives. This 
assumption is satisfied if there are no plasma instabilities in the 
reconnection layer, and the reconnection rate is sub-Alfvenic, 
$E_z\ll V_AB_{ext}$. Here $V_A\equiv B_{ext}/\sqrt{nm_p}$ is the 
Alfven velocity. Third, the pressure tensors are isotropic, so that 
the pressure terms in Eqs.~(\ref{OHMS_LAW}) and~(\ref{MOTION_LAW}) 
are scalars. 

Using Ampere's law and neglecting the displacement current, 
we find the current components to be 
$j_x=\partial_y B_z$, $j_y=-\partial_x B_z$ and 
$j_z=\partial_x B_y-\partial_y B_x$. 
The $z$-component of the current at the central point~$O$ is
\begin{eqnarray}
j_o\equiv (j_z)_o=
\left(\partial_x B_y-\partial_y B_x\right)_o
\approx B_{ext}/\delta,
\label{AMPERES_LAW}
\end{eqnarray}
where we use the estimates 
$(\partial_y B_x)_o\ll (\partial_x B_y)_o$ and 
$(\partial_x B_y)_o\approx B_{ext}/\delta$ 
at the point~$O$. The last estimate follows directly 
from the definition of $\delta$.

Equation~(\ref{MOTION_LAW}) for the plasma (proton) acceleration 
along the reconnection layer gives
\begin{eqnarray}
nm_p({\bf V}{\bf\nabla})V_y+d_e^2({\bf j}{\bf\nabla})j_y
=-\partial_y P+j_zB_x-j_xB_z.
\label{MOTION_LAW_Y}
\end{eqnarray}
The $y$ derivative of this equation at the point~$O$ gives
\begin{eqnarray}
nm_p(\partial_y V_y)_o^{\,2}(1+d_e^2{\tilde\gamma}^2/d_p^2)
\approx 2B_{ext}^2/L^2+j_o(\partial_y B_x)_o.
\label{ACCELERATION}
\end{eqnarray}
Here we introduce a useful dimensional parameter 
\begin{eqnarray}
{\tilde\gamma}\equiv 
(\partial_{xy}B_z)_o\left/ne(\partial_y V_y)_o\right.,
\label{GAMMA_TILDE}
\end{eqnarray}
which measures the relative strength of the Hall term 
$({\bf j}\!\times\!{\bf B})_z/ne$ and the ideal MHD term 
$({\bf V}\!\times\!{\bf B})_z$ inside the electron layer. 
In the derivation of Eq.~(\ref{ACCELERATION}) we use the estimate 
$(\partial_{yy}P)_o\approx(\partial_{yy}B_y^2/2)_{ext}\approx-2B_{ext}^2/L^2$, 
which follows from the force balance condition for the slowly 
inflowing plasma across the layer~\citep{malyshkin_2005}. 

Faraday's law 
${{\bf\nabla}\times{\bf E}}=-\partial_t{\bf B}$ for
the $x$- and $y$-components of a quasi-stationary magnetic field 
in two dimensions gives
$\partial_y E_z=-\partial_t B_x\approx0$
and $\partial_x E_z=\partial_t B_y\approx0$. 
Thus, $E_z$ is approximately 
constant in space, and from the generalized Ohm's 
law~(\ref{OHMS_LAW}) we obtain
\begin{eqnarray}
E_z &\!=\!& \eta j_z-V_xB_y+V_yB_x+(j_xB_y-j_yB_x)/ne
\nonumber\\
&\!+\!& d_e^2\left[j_x\partial_xV_z+j_y\partial_yV_z+V_x\partial_xj_z+V_y\partial_yj_z\right.
\nonumber\\
&&\quad\;\left.-(j_x\partial_xj_z+j_y\partial_yj_z)/ne\right]
\approx\mbox{constant}.
\quad
\label{OHMS_LAW_Z}
\end{eqnarray}
The reconnection rate is determined by the value of $E_z$ 
at the point~$O$, namely $E_z=\eta j_o$. 
We estimate $j_o$ below.

Taking the second derivatives of the z-component of 
Eq.~(\ref{MOTION_LAW}) with respect to $x$ and $y$ at the point~$O$, 
we find 
$(\partial_{xx}V_z)_o=-(e{\tilde\gamma}/m_p)[(\partial_x B_y)_o-d_e^2(\partial_{xx}j_z)_o]$
and 
$(\partial_{yy}V_z)_o= (e{\tilde\gamma}/m_p)[(\partial_y B_x)_o+d_e^2(\partial_{yy}j_z)_o]$.
Using these expressions, we calculate the second derivatives 
of Eq.~(\ref{OHMS_LAW_Z}) with respect to $x$ and $y$ at the point~$O$ 
and obtain
\begin{eqnarray}
\eta j_o/\delta^2 &\!\approx\!&
2(\partial_y V_y)_oj_o[1+d_e^2/\delta^2]
\nonumber\\
&\!\!& \times\big[1+{\tilde\gamma}(1-d_e^2{\tilde\gamma}/d_p^2)\big],
\label{E_Z_PERPENDICULAR}
\\
\eta j_o/L^2 &\!\approx\!& 
2(\partial_y V_y)_o[(\partial_y B_x)_o-d_e^2j_o/L^2]
\nonumber\\
&\!\!& \times\big[1+{\tilde\gamma}(1-d_e^2{\tilde\gamma}/d_p^2)\big],
\label{E_Z_PARALLEL}
\end{eqnarray}
where we use $(\partial_x V_x)_o=-(\partial_y V_y)_o$,
$(\partial_x B_y)_o\approx j_o$, $(\partial_{xx}j_z)_o\approx-j_o/\delta^2$ 
and $(\partial_{yy}j_z)_o\approx-j_o/L^2$.

In Eq.~(\ref{OHMS_LAW_Z}), the electric field $E_z$ is balanced by the 
MHD and Hall terms outside the electron layer, where the electron inertia 
terms are unimportant. Therefore, 
\begin{eqnarray}
E_z &\!\!\approx\!\!& -V_xB_y(1\!-\!j_x/neV_x)\approx 
(\partial_y V_y)_o\delta\, B_{ext}(1\!+\!{\tilde\gamma}),
\label{E_Z_X=DELTA}
\\
E_z &\!\!\approx\!\!& V_yB_x(1\!-\!j_y/neV_y)\approx 
(\partial_y V_y)_o(\partial_y B_x)_oL^2(1\!+\!{\tilde\gamma})
\qquad
\label{E_Z_Y=L}
\end{eqnarray}
at the points $(x\!\approx\!\delta, y\!=\!0)$ and $(x\!=\!0,y\!\approx\!L)$ 
respectively. Here we use the estimates
$j_x\approx(\partial_{xy}B_z)_o\delta$, $j_y\approx-(\partial_{xy}B_z)_oL$, 
$V_x\approx-(\partial_y V_y)_o\delta$, $V_y\approx(\partial_y V_y)_oL$,
$B_x\approx (\partial_y B_x)_oL$ and $B_y\approx B_{ext}$.

The ratio of Eqs.~(\ref{E_Z_PERPENDICULAR}) and~(\ref{E_Z_PARALLEL})
gives $(\partial_y B_x)_o\approx(\delta^2j_o/L^2)(1+2d_e^2/\delta^2)$, 
while the ratio of Eqs.~(\ref{E_Z_X=DELTA}) and~(\ref{E_Z_Y=L}) gives 
$(\partial_y B_x)_o\approx B_{ext}\delta/L^2\approx\delta^2j_o/L^2$,
where we use Eq.~(\ref{AMPERES_LAW}). Comparing these two estimates, 
we find $\delta\gtrsim d_e$; therefore, $j_o\lesssim B_{ext}/d_e$
and $E_z\lesssim \eta B_{ext}/d_e$ \citep{zocco_2008}.

Next, we use the $z$-component of Faraday's law:
$0\approx-\partial_t B_z=\partial_x E_y-\partial_y E_x$.
Taking the $\partial_{xy}$ derivative of this equation at the 
point~$O$, and using Eq.~(\ref{OHMS_LAW}) for 
$E_x$ and $E_y$, 
after tedious but straightforward derivations, 
we obtain
\begin{eqnarray}
0 &\!\approx\!&
-\eta\left[(\partial_{xyxx}B_z)_o+(\partial_{xyyy}B_z)_o\right]
+(1-d_e^2{\tilde\gamma}/d_p^2)
\nonumber\\
&&\times[(\partial_y B_x)_o(\partial_{xx}j_z)_o
+(\partial_x B_y)_o(\partial_{yy}j_z)_o]/ne
\nonumber\\
&\!\approx\!& 
\eta ne(\partial_y V_y)_o{\tilde\gamma}/\delta^2
-(1-d_e^2{\tilde\gamma}/d_p^2)
\nonumber\\
&&\times[(\partial_y B_x)_oj_o/\delta^2+j_o^2/L^2]/ne.
\quad
\label{E_XY}
\end{eqnarray}
To derive the final expression, we use Eq.~(\ref{GAMMA_TILDE}) 
and the estimates
$(\partial_{xyxx}B_z)_o\approx-(\partial_{xy}B_z)_o/\delta^2
\gg(\partial_{xyyy}B_z)_o$,
$(\partial_{xx}j_z)_o\approx-j_o/\delta^2$,
$(\partial_{yy}j_z)_o\approx-j_o/L^2$, 
$(\partial_x B_y)_o\approx j_o$. Note that Eq.~(\ref{E_XY}) 
results in ${\tilde\gamma}<d_p^2/d_e^2$.

We estimate the proton layer half-thickness $\Delta$ as follows. 
Outside the electron layer the electron inertia and magnetic tension 
terms can be neglected in Eq.~(\ref{MOTION_LAW_Y}), and we have
$nm_p({\bf V}{\bf\nabla})V_y\approx-\partial_y P$. Taking
the $y$ derivative of this equation at $y\!=\!0$, we obtain 
$nm_p[V_x(\partial_{xy} V_y)+(\partial_y V_y)^{\,2}]
\approx-(\partial_{yy}P)_o\approx 2B_{ext}^2/L^2$. Here the
term $V_x(\partial_{xy} V_y)$ is about of the same size as 
the term $(\partial_y V_y)^{\,2}$. Therefore, 
we find that $(\partial_y V_y)_{ext}\approx V_A/L$ 
outside the electron layer (but inside the proton layer). 
Next, in the upstream region outside the proton layer
ideal single-fluid MHD applies. As a result, at 
$x\approx\Delta$ and $y=0$ Eq.~(\ref{OHMS_LAW_Z}) reduces to 
$E_z=\eta j_o\approx -V_xB_y\approx(\partial_y V_y)_{ext}\Delta B_{ext}
\approx \Delta B_{ext}V_A/L$. Thus, 
\begin{eqnarray}
(\partial_y V_y)_{ext}\approx V_A/L,
\qquad
\Delta\approx \eta j_oL/V_AB_{ext}.
\label{DELTA}
\end{eqnarray}

Now we solve equations~(\ref{AMPERES_LAW}), 
(\ref{ACCELERATION})--(\ref{GAMMA_TILDE}), 
(\ref{E_Z_PERPENDICULAR})--(\ref{DELTA}) for unknown quantities 
$j_o$, $\delta$, $\Delta$, $L$, ${\tilde\gamma}$, $(\partial_y V_y)_o$, 
$(\partial_y B_x)_o$ and $(\partial_{xy}B_z)_o$. We neglect 
factors of order unity, and we treat the external field $B_{ext}$ 
and scale $L_{ext}$ as model parameters. 
Recall that parameter ${\tilde\gamma}$, given by Eq.~(\ref{GAMMA_TILDE}), 
measures the relative strength of the Hall term and the ideal MHD term 
in the z-component of the Ohm's law. Depending on the
value of ${\tilde\gamma}$, we have the following solutions and 
the corresponding reconnection regimes.

{\bf Sweet-Parker reconnection.} 
When ${\tilde\gamma}\lesssim 1$, both the Hall current and electron 
inertia are negligible, and the electrons and protons 
flow together. In this case, we obtain the Sweet-Parker solution: 
$j_o\approx \sqrt{S}\,B_{ext}/L_{ext}$,
$E_z=\eta j_o\approx V_AB_{ext}/\sqrt{S}$,
$\delta\approx \Delta\approx L_{ext}/\sqrt{S}$,
$L\approx L_{ext}$, 
${\tilde\gamma}\approx Sd_p^2/L_{ext}^2$,
$(\partial_y V_y)_o\approx(\partial_y V_y)_{ext}\approx V_A/L_{ext}$,
$(\partial_y B_x)_o\approx B_{ext}/L_{ext}\sqrt{S}$,
$(\partial_{xy}B_z)_o\approx SB_{ext}d_p/L_{ext}^3$,
where $S\equiv V_AL_{ext}/\eta$ is the Lundquist number.
Condition ${\tilde\gamma}\lesssim 1$
gives $S\lesssim L_{ext}^2/d_p^2$. Therefore, Sweet-Parker 
reconnection occurs when $d_p$ is less than the Sweet-Parker layer thickness,  
$d_p\lesssim L_{ext}/\sqrt{S}$~\citep{drake_2006,yamada_2009,zweibel_2009}.

{\bf Hall reconnection.} 
When $1\!\lesssim\!{\tilde\gamma}\!\lesssim\! d_p/d_e$, 
the Hall current is important but the electron inertia is negligible. 
In this case, the solution is 
$j_o\approx B_{ext}^2/\eta neL = Sd_pB_{ext}/LL_{ext}$,
$E_z\approx B_{ext}^2/neL=(d_p/L)V_AB_{ext}$,
$\delta\approx ne\eta L/B_{ext}=LL_{ext}/Sd_p$,
$\Delta\approx d_p$,
${\tilde\gamma}\approx Sd_p^2/LL_{ext}$,
$(\partial_y V_y)_o\approx (\partial_y V_y)_{ext}\approx V_A/L$,
$(\partial_y B_x)_o\approx ne\eta/L=B_{ext}L_{ext}/Sd_pL$ and
$(\partial_{xy} B_z)_o\approx B_{ext}^2/ne\eta L^2
=Sd_pB_{ext}/L^2L_{ext}$. These results are in agreement 
with previous theoretical findings~\citep{cowley_1985,bhattacharjee_2001,
simakov_2008,malyshkin_2008}.
Condition $1\lesssim {\tilde\gamma}\lesssim d_p/d_e$
translates into $Sd_ed_p/L_{ext}\lesssim L\lesssim Sd_p^2/L_{ext}$
for the electron layer length $L$. Unfortunately, the exact value of 
$L$ cannot be estimated from Eqs.~(\ref{AMPERES_LAW})--(\ref{DELTA}) 
in the Hall reconnection regime. 
In studies~\citep{cowley_1985,simakov_2008,malyshkin_2008} 
$L$ was essentially treated as a fixed parameter. 
Here, we take a different approach and make a conjecture that 
the Hall reconnection regime describes a transition from the slow 
Sweet-Parker reconnection to the fast collisionless reconnection 
(presented below). Numerical simulations and experiment have 
demonstrated that this transition occurs when 
$d_p\approx L_{ext}/\sqrt{S}$~\citep{huba_2004,drake_2006,
yamada_2009,zweibel_2009}. 
Therefore, our conjecture implies that the Hall reconnection solution is
$S\approx L_{ext}^2/d_p^2$,
$L_{ext} \gtrsim L \gtrsim d_eL_{ext}/d_p$,
$j_o\approx B_{ext}L_{ext}/d_pL$,
$E_z\approx(d_p/L)V_AB_{ext}$,
$\delta\approx d_pL/L_{ext}$,
$\Delta\approx d_p$,
${\tilde\gamma}\approx L_{ext}/L$,
$(\partial_y V_y)_o\approx(\partial_y V_y)_{ext}\approx V_A/L$,
$(\partial_y B_x)_o\approx B_{ext}d_p/LL_{ext}$ and
$(\partial_{xy} B_z)_o\approx B_{ext}L_{ext}/d_pL^2$.
As the electron layer length $L$ decreases from its maximal
value $L\approx L_{ext}$ to its minimal value 
$L\approx d_eL_{ext}/d_p$, this solution changes from 
the slow Sweet-Parker solution to the fast collisionless reconnection 
solution that is presented next. 

{\bf Collisionless reconnection.} 
When $d_p/d_e\lesssim{\tilde\gamma}<d_p^2/d_e^2$, the electron 
inertia and the Hall current are important inside the electron 
layer and the proton layer respectively. In this case, 
the solution is 
\begin{eqnarray}
j_o &\!\approx\!& B_{ext}/d_e,
\qquad 
\delta\approx d_e,
\label{J_O_FAST}
\\
E_z\!=\eta j_o &\!\approx\!& \eta B_{ext}/d_e=(L_{ext}/Sd_e)V_AB_{ext}
\nonumber\\
&\!\approx\!& (\Delta/L)V_AB_{ext}\approx (d_p/L)V_AB_{ext},
\label{Ez_FAST}
\\
L &\!\approx\!& V_Ad_ed_p/\eta = Sd_ed_p/L_{ext},
\label{L_FAST}
\\
d_p/d_e \lesssim {\tilde\gamma} &\!<\!& d_p^2/d_e^2,
\qquad\quad
\Delta \approx d_p,
\label{GAMMA_FAST}
\\
(\partial_y V_y)_o &\!\approx\!& \eta/d_e^2{\tilde\gamma}
=V_AL_{ext}/Sd_e^2{\tilde\gamma},
\label{Vy_Y_FAST}
\\
(\partial_y V_y)_{ext} &\!\approx\!& \eta/d_ed_p=V_AL_{ext}/Sd_ed_p\approx V_A/L,
\label{Vy_Y_EXT_FAST}
\\
(\partial_y B_x)_o &\!\approx\!& B_{ext}\eta^2/V_A^2d_ed_p^2
=B_{ext}L_{ext}^2/S^2d_ed_p^2,\quad
\label{Bx_Y_FAST}
\\
(\partial_{xy} B_z)_o &\!\approx\!& B_{ext}\eta/V_Ad_e^2d_p
=B_{ext}L_{ext}/Sd_e^2d_p.
\label{Bz_XY_FAST}
\end{eqnarray}
Apart from the definition of the reconnecting field $B_{ext}$, 
Eqs.~(\ref{J_O_FAST})--(\ref{L_FAST}) essentially coincide with the 
results obtained in~\citep{zocco_2008} for an electron MHD (EMHD) 
reconnection model.
Note that the value of ${\tilde\gamma}$ or, alternatively, the 
value of the proton acceleration rate $(\partial_y V_y)_o$ at the 
point~$O$ cannot be determined exactly. This is because 
in the plasma motion equation~(\ref{MOTION_LAW_Y}), the magnetic tension 
and pressure forces are balanced by the electron inertia term 
$d_e^2({\bf j}{\bf\nabla})j_y$ inside the electron layer.
The proton inertia term $nm_p({\bf V}{\bf\nabla})V_y$ can be of the 
same order or smaller, resulting in the upper limit 
$(\partial_y V_y)_o\lesssim V_A/L$. 
Thus, inside the electron layer the magnetic 
energy is converted into the kinetic energy of the electrons 
(and into Ohmic heat), while the proton kinetic energy can be much 
smaller. However, in the downstream region $y\gtrsim L$, as the 
electrons gradually decelerate, their kinetic energy is converted
into the proton kinetic energy. As a result, the eventual proton 
outflow velocity becomes $\approx\!V_A$~\footnote{
This result can be obtained by integration of Eq.~(\ref{MOTION_LAW_Y})
over $y$; the term $d_e^2({\bf j}{\bf\nabla})j_y$ integrates 
to zero.}.
These results emphasize the critical role that electron inertia 
plays in the plasma momentum equation~(\ref{MOTION_LAW}). 

The collisionless reconnection rate given by Eq.~(\ref{Ez_FAST}), 
although being proportional to resistivity~\footnote{
We use the standard term ``collisionless'' for this regime because in 
this regime $\eta$ is the effective resistivity, which is to be 
calculated from kinetic theory.
}, 
is much faster than the Sweet-Parker rate 
$E_z\approx V_AB_{ext}/\sqrt{S}$ as long as $S\ll L_{ext}^2/d_e^2$.
The solution~(\ref{J_O_FAST})--(\ref{Bz_XY_FAST}) is valid
provided $L_{ext}/d_e\ll S\lesssim L_{ext}^2/d_ed_p$, which is obtained
from conditions $E_z\ll V_AB_{ext}$ and $L\lesssim L_{ext}$. Thus,
both fast collisionless and slow Sweet-Parker reconnection regimes
can exist simultaneously, as found in simulations~\citep{cassak_2005}. 
These simulations also found a hysteresis for transition between slow 
and fast reconnection regimes. This implies that the 
transition Hall regime, during which the electron layer length $L$ 
decreases, may occur at Lundquist numbers other than 
$S\approx L_{ext}^2/d_p^2$, depending on the past history. 
Unfortunately, our stationary model cannot describe time-dependent 
transition processes.
 
It is known that the single-fluid MHD reconnection becomes much faster 
when resistivity $\eta$ is anomalously enhanced by current-driven plasma 
instabilities~\citep{kulsrud_2001,malyshkin_2005,yamada_2009,zweibel_2009}. 
Eq.~(\ref{Ez_FAST}) shows that resistivity enhancement can 
considerably increase the collisionless reconnection rate as well.
This enhancement can occur after the electric current value ($j_o$) 
jumps up during the transition from the Sweet-Parker to the collisionless 
regime at $d_p\approx L_{ext}/\sqrt{S}$, and could be a physical 
mechanism of very fast reconnection.


\section{\label{DISCUSSION}
Discussion
}
Let us compare theoretical results~(\ref{J_O_FAST})--(\ref{Bz_XY_FAST}) 
for collisionless reconnection with 
numerical simulations and experiment.
The estimates $\Delta\approx d_p$ for the proton layer thickness, 
$\delta\approx d_e$ for the electron layer thickness, 
$B_z\approx (\partial_{xy} B_z)_o\delta L\approx B_{ext}$ for
the quadrupole field, and 
$u_y^e\approx-j_y/ne\approx (\partial_{xy} B_z)_oL/ne
\approx V_{eA}\equiv B_{ext}/\sqrt{nm_e}$ for the electron outflow 
velocity agree with 
simulations~\citep{drake_2006,yamada_2009,zweibel_2009,daughton_2006,
fujimoto_2006,daughton_2007,shay_2007}.
The estimates $\Delta\approx d_p$ and $B_z\approx B_{ext}$ also 
agree with experiment~\citep{yamada_2009}. However, 
the experimentally measured electron layer thickness is about eight 
times larger than the model and simulations 
predict~\citep{ren_yamada_2008,ji_daughton_2008}. 
Three-dimensional geometry effects and plasma instabilities 
make direct comparison and exact interpretation of the experimental 
results difficult~\citep{yamada_2009,ji_daughton_2008}. 

Our theoretical results are qualitatively consistent with recent 
numerical findings of an inner electron  dissipation layer and 
of electron outflow jets that extend into the proton 
layer~\citep{daughton_2006,fujimoto_2006,daughton_2007,shay_2007}. 
The electron layer length $L\propto d_e\propto m_e^{1/2}$ decreases
with the electron mass, as in simulations~\citep{shay_2007}, but
the scaling law observed in these simulations was slightly different, 
$L\propto m_e^{3/8}$. Length $L\approx Sd_ed_p/L_{ext}$ is generally
much larger than both $\delta\approx d_e$ and $\Delta\approx d_p$, 
consistent with simulations~\citep{daughton_2006,fujimoto_2006,daughton_2007}.
However, if resistivity $\eta$ becomes anomalous and enhanced over 
the Spitzer value so much that $S\approx L_{ext}/d_e$, then $L$ can 
theoretically become of order of $d_p$, as in 
simulations~\citep{huba_2004,shay_2007}.
Our theoretical results for the proton velocity ${\bf V}$ agree  
with simulations~\citep{daughton_2007}, which found the proton 
outflow velocity to be significantly less than $V_A$ and also found 
acceleration of protons in the decelerating electron jets. 
Unfortunately, detailed quantitative comparison of our results to the 
results of kinetic numerical simulations is hindered because the 
simulations do not explicitly specify resistivity $\eta$. Also, in the 
simulations the electron pressure tensor anisotropy was found to play 
a critical role inside the electron layer and 
jets~\citep{daughton_2007,shay_2007}, while in this study an isotropic 
pressure is assumed and the electrons are coupled to the field 
lines inside the electron outflow jets.
Thus, in our model the electric field $E_z$ is supported by the Hall 
term $({\bf j}\times{\bf B})_z/ne$ in the downstream region $y\gtrsim L$.
As a result, there are Hall-MHD Petschek shocks attached to the 
ends of the electron layer~\footnote{
The magnetic field and electron velocity parallel components jump across 
the shocks. The opening angle between the shocks is 
$\alpha\approx B_x/B_y\approx L_{ext}/Sd_p\ll 1$. 
}, as observed in numerical simulations~\citep{arber_2006}. However, 
in these simulations a spatially localized anomalous resistivity was 
prescribed, resulting in a short layer length, while here
resistivity $\eta$ is assumed to be constant.

I would like to thank F.~Cattaneo, H.~Ji, D.~Lecoanet, 
R.~Kulsrud, J.~Mason, A.~Obabko, D.~Uzdensky 
and M.~Yamada for very useful discussions. 
This study was supported by the NSF Center for Magnetic 
Self-Organization (CMSO), NSF award \#PHY-0821899.



\end{document}